\documentclass[5p,times]{elsarticle}
\usepackage{epsfig}
\usepackage{graphicx}
\usepackage{amssymb}
\usepackage[figuresright]{rotating}
\usepackage{amssymb}
\def\bea {\begin{eqnarray}}
\def\eea {\end{eqnarray}}

\def\be {\begin{equation}}
\def\ee {\end{equation}}

\begin{document}
\begin{frontmatter}

\title{ Cosmic Separation of Phases: the microsecond universe and the 
neutron star.}

\author{\it Bikash Sinha $^{1,2}$}
\ead{bikash@vecc.gov.in}

\address{$^1$ Bose Institute, 93/1 APC Road, Kolkata-700009, India \\
$^2$ Variable Energy Cyclotron 
Centre, 1/AF Bidhan Nagar, Kolkata-700064, India.}            

\begin{abstract}
It is entirely plausible under reasonable condition, that a first 
order QCD phase transition occurred from quarks to hadrons when the 
universe was about a microsecond old. Relics, if there be any, after 
the quark hadron phase transition are the most deciding signatures 
of the phase transition. It is shown in this paper that the 
quark nuggets, possible relics of first order QCD phase transitions 
with baryon number larger than $10^{43}$ will survive the entire 
history of the universe uptil now and can be considered as candidates 
for the cold dark matter. The spin down core of the neutron star on 
the high density low temperature end of the phase diagramme initiates 
transition from hadrons to quarks. As the star spins down, the size of 
the core goes on increasing. Recently discovered massive Pulsar PSRJ 
1614-2230 with a mass of 1.97$\pm$0.04 $M_{\odot}$ most likely has a 
strongly interacting quark core. What possible observables can there 
be from these neutron stars? 

\end{abstract}

%\pacs{13.85.-t,25.75.Nq}
%\pacs{25.75.-q,25.75.Dw,24.85.+p}
\end{frontmatter}
\section{Introduction}
\label{}
It is by now conventional wisdom, that collisions between two nuclei at
relativistic energies such as at RHIC as well as LHC, lead to the 
formation of quark gluon plasma (QGP). A large number of novel and 
exciting discoveries have been made \cite{qm12pro,qm08pro}. One of the most 
surprising observation\cite{npa05} is to do with the discovery that 
QGP formed at RHIC and now at LHC does not behave as a non interacting 
gas of quarks and gluons, as anticipated theoretically but as a "perfect 
fluid" of very small value of shear viscosity to its entropy density. 
The deep implication of this recent observation in the cosmological 
arena of microsecond old universe and the neutron star will be 
discussed. 
%Consequent of the same conventional wisdom it is expected 
%that the universe, a microsecond old, goes through a phase transition 
%from QGP to hadrons. 

Since the chemical potential of the universe, at that primordial epoch 
was close to zero and the temperature was around 150 MeV
~\cite{hotqcd,bielefield}, 
the wisdom of lattice will lead the universe to crossover from the 
universe of quarks and gluons to a universe of hadrons, erasing all the 
memories, with no relic of the earlier phase of quarks and gluons. 
It should be noted at this point that the lattice calculation, which 
is static, is hardly applicable to an expanding universe at a non zero 
temperature and radiation dominated with bare quark masses.
%  a The universe on the other hand 
%when about a micro second old goes through a phase transition from 
%QGP to hadrons. Since the chemical potential of the universe, at that 
%primordial epoch was close to zero and the temperature was around 
%150 MeV~\cite{hotqcd,bielefield}, the wisdom of lattice will lead 
%the universe to cross over to hadrons, erasing all the memories of 
%the universe at earlier epochs with virtually no relic of the QCD 
%phase transition.

Sometime ago Witten ~\cite{witten84} argued for 
a first order phase transition with "small" supercooling which means 
that the transition effectively occurs at a temperature at which most 
of the latent heat between QGP and hadrons still remains, so that the 
phase co-existence can be established after nucleation. 

%there are several compelling evidences that QGP is formed
%~\cite{brahmswhitepaper,phoboswhitepaper,starwhitepaper,phenixwhitepaper} at RHIC energies. 
\par
More recently Boeckel and Schaffner-Bielich~\cite{boeckelprd12} by 
introducing "little inflation" scenario at the point of quark hadron 
phase transition have demonstrated that the universe goes through 
a first order phase transition without contradicting any contemporary 
cosmological observation. 

\section{The Microsecond Universe}
At this point it is highly instructive to recapitulate the raiseon 
d'$\hat{\mathrm e}$tre of little inflation and its relevance to that primordial 
epoch of quark hadron phase transition.

It is to do with baryogenesis. One of the more compelling scenarios 
of baryogenesis is based on its generation from leptogenesis through 
topological sphaleron transitions occuring around the electroweak 
transition temperature. Leptogenesis, in its turn, appears through 
out-of-equilibrium decays of heavy right-handed neutrinos which occur 
naturally via a {\it seasaw} mechanism leading to Majorana masses  
for neutrinos (as well as neutrino oscillation parameters) within 
observable ranges. It is supposedly Majorana nature of neutrinos 
which lies at the heart of the incipient lepton number violation. 
If, on the otherhand, neutrinoless double beta decay experiments 
yield null results, and neutrinos are confirmed to be Dirac fermions, 
the scenario of baryogenesis loses its prime attraction, entailing 
unsavoury fine tuning. 

Given such volatile situation, alternative scenarios of baryogenesis 
can not be ruled out. Prominent among these is the non-thermal 
Affleck-Dine ~\cite{affleck85} mechanism based upon out-of-equilibrium 
decays of heavy quarks and leptons (which respectively carry 
baryon and lepton number) within a super symmetric framework. When 
supersymmetry is unbroken, the scalar potential for quarks and leptons 
has flat directions which permit these scalars to acquire very large 
vacuum value.

The Affleck-Dine mechanism has the potential to produce a baryon assymetry 
of O(1) without requiring superhigh temperatures. However, the observed 
baryon assymetry of O($10^{-10}$) at CMB temperatures needs to emerge 
naturally from such a scenario. This is what is achieved through a 
'little inflation' of about 7 e-foldings occuring at a lower temperature 
which may be identified with the QCD phase transition thought of as 
a first order phase transition. The universe is thus assumed to begin 
with a large baryon chemical potential acquired through an Affleck-Dine 
type of mechanism and then undergoes a period of inflation after crossing the 
first order QCD phase transition line, while remaining in a deconfined and 
in a chirally symmetric phase. The delayed phase transition then releases 
the latent heat and produces concomittantly a large entropy density 
which reduces the baryon assymetry to currently observed values. It then 
enters a reheating phase all the way up to the usual reheating 
 temperature with no significant change in the baryon potential and then 
follows the standard path to lower temperature.
%%%%%%%%%%%%%%%%%%%%%%%%%%%%%%%%%%%%%%%%%%%%%%%%%%%%%%%%%%%%%%%%%%%%%
\par
The universe is then trapped in a false metastable QCD 
vacuum state. A delayed phase transition takes place. The latent heat thus 
released caused a large entropy release, diluting the baryon asymmetry to 
the presently observed value. The microsecond old universe goes 
from the universe of quarks to universe of hadrons in a first order phase 
transition. The question one naturally goes on to ask " what possible 
relic(s) of that primordial epoch can we observe today?"

It is clear however that existence of relics, (say) in the form of Quark 
Nuggets is a definitive signature of a first order phase transition. 
Thus one such relics are conclusively discovered, all debates should 
converge. 

\par
Witten pointed out in that seminal paper ~\cite{witten84} that consequent 
to the first order phase transition stable "quark nuggets (QN)" made of  
"strange quark matter" may indeed survive till the present epoch. These 
nuggets may well be good candidates for baryonic dark 
matter in the universe ~\cite{raha05,alam99,bhattacharya00}. The strange nuggets 
may also provide closure density without violating the basic 
premise of neucleosynthesis ~\cite{alam99,bhattacharya00}.

A QN formed after the phase transition will survive up to the present 
epoch if it looses heat and cools without loosing any significant fraction 
of initial baryon number.

On the basis of simple heuristic argument a QN with baryon number $N_B$ 
at a time $t$ will stop evaporating further and thus survive forever, 
that means, now, if the time scale of evaporation 
$\tau_{ev} (N_B,t)$ =$N_B/(dN_B/dt)$ is considerably larger than the 
Hubble expansion time scale, $H^{-1}(t)=2t$, of the universe. Using 
Chromo Electric flux tube model $N_B$ and $dN_B/dt$ 
was calculated in ref~\cite{bhattacharjee93}. Taking the Hubble constant 
$H_0$=75 km ~sec$^{-1}$~Mpc$^{-1}$ it was shown that with 
$N_{B,in} \le 10^{43.25}$ and a flat universe, the nuggets will be 
unstable and evaporate completely with time. In contrast QN's with 
$N_{B,in} \ge 10^{43.25}$, with time do not loose any baryon and 
thus survive for ever. $N_B$ considerably low 
than $10^{43.25}$ (say) $10^{42}$ evaporate very early, 
only a few microsecond after the phase transition from 
quarks to hadrons. The neutron to proton ratio in the ambient 
universe, evidently goes up due to the evaporation (preferably) 
neutrons from the surface of quark nuggets. Protons, on the other 
hand tend to get coulomb repelled by the ambient cosmic soup. 
Such differential increase or overdensity would dissipate away, 
primarily because of the conduction of heat into the baryon 
overdense region by the neutrinos coming from the ambient universe. 
This process goes on until the temperature is of the 
order of 1 MeV. After that epoch baryon diffusion  
dominates the scenario\cite{raha05,alam99}.   
\par
As per standard big bang primordial (SBBN) nucleosynthesis, the baryons 
constitute only $\sim 10 \%$ of the closure density 
($\Omega_B \sim 0.1)$. Total baryon number of $10^{50}$ within the 
horizon at a temperature of $\sim$ 100 MeV \cite{raha05} would close the universe 
baryonically provided these baryons do not take part in SBBN, a 
criteria ideally fulfilled by the QN's. This requires 
$N_B^{QN} \le 10^{45.3}$ which is clearly above the survivability limit 
of QN's as mentioned earlier \cite{bhattacharjee93}. 
%%%%%%%%%%%%%%%%%%%%%%%%%%%%%%%%%%%%%%%%%%%%%%%%%%%%%%%%%%%%%%%%%%%%%
%\begin{figure}
%\begin{center}
%\includegraphics[scale=0.43]{fig_direct_rhicphoton_2.eps}
%\caption{The $p_T$ spectra of direct photons from Au+Au collisions
%at $\sqrt{s_{NN}}$=200 GeV RHIC energy. The solid circles are data points 
%for 0-5\% centrality measured by PHENIX collaboration~\cite{phenix12}. 
%The theoretical evaluations from different sources are displayed.}
%\label{fig1}
%\end{center}
%\end{figure}
%%%%%%%%%%%%%%%%%%%%%%%%%%%%%%%%%%%%%%%%%%%%%%%%%%%%%%%%%%%%%%%%%%%%%
\par
Thus, one can conclude at this stage that the baryons contained in QN's 
will not participate in nucleosynthesis. In 
the galaxy formation they would behave like planetary mass black holes.

It is also to be noted that in equilibrium the energy per quark equals 
the chemical potential, so that the energy per quark in strange quark matter 
is less than the energy per quark in zero strangeness quark matter; the 
strange quark matter is more tightly bound than non strange quark matter 
by about 100 MeV per baryon. Thus energetically it is more convenient 
to have QN's made of strange quarks (SQN).

%Now what happens to the SQN's? It has already been argued ~\cite{bhattacharya00} that 
%the entire cold dark matter (CDM) ($\Omega_{CDM} \sim 0.3-0.5$) can be 
%explained by SQN's.  

What about the number density of SQN's? 

It is seen ~\cite{raha05,alam99,bhattacharya00} that the 
number density of QN's, $n_{QN}$ is given by 

\begin{equation}
n_{QN} \approx \frac{1968}{({\mathrm v}t_P)^3} 
\label{eq1}
\end{equation}

%where we notice that if $\lambda$ is small, (say) 0.01 only 1$\%$ of 
%high temperature domains will become QN's; 
$t_P$ is typically the 
percolation time and $t_P \equiv 27 \mu s$ ~\cite{alam99}. 
%Insisting that 
%the universe is closed by the baryonic dark matter trapped inside QN's, we have
In an idealized situation where the universe is closed by the baryonic dark 
matter trapped in the QN's we have  
\begin{equation}
{N_B}^H (t_P)=n_{QN} V_H (t_P) {N_B}{QN}
\label{eq2}
\end{equation}

where ${N_B}^H(t_P)$ is the total number of baryons, required to close 
the universe ($\Omega_B =1$) at $t_P$, ${N_B}^{QN}$ is the total number 
of baryons contained in a single quark nugget and $V_H (t_P)$ is the 
horizon volume ($V_H (t_P)=(4\pi/3)(ct_P)^3$). Taking ${\mathrm v}/c=1{\sqrt{3}}$ 
we get ${N_B}^{QN} \le 10^{-4.7} {N_B}^H (t_P)$. Using Eqs.~\ref{eq1} and 
\ref{eq2} the conservative upper limit of the baryon number of individual 
QN's thus is comfortably below the number $10^{49}$, so that the sizable 
number of QN's can be formed. Their size distribution peaks for reasonable 
nucleation rates at baryon number $\sim 10^{42-44}$ \cite{alam99,bhattacharya00} evidently 
stable and much lower than the horizon limit. It is also found 
~\cite{witten84} that SQN's contain 80-90$\%$ of the mass of the baryons 
of the universe, hinting that the large mass reservoir in the universe 
are contained in SQN's. 

It appears that the upper limit on the baryon number of QN's that would 
close the universe ~\cite{alam99} is not very sensitive to the nucleation 
mechanism and estimates point to the real possibility that SQN's indeed 
are possible candidate of cold dark matter and even can close the universe.

Initially as is well known, the radiation pressure will keep the SQN's 
from clumping under gravity. Once, with time the gravitational force 
starts dominating, SQN's will tend to coalesce under mutual gravity. 
Detailed calculation ~\cite{bhattacharya00} for baryon number $b_N$ at the critical 
temperature $T_{cl}$ (MeV), the total number $N_{QN}$ of SQN's are compared 
scaled and scaled by $M_{\odot}$ in Table ~\ref{tbl1}.
With increase in $b_N$, the number $N_{QN}$ drops rapidly.
{\footnote{ Most significant and realistic contribution comes from $b_N \equiv 10^{42}$, 
for $10^{46}$, $M/M_{\odot}$ is already too small; for $b_N \equiv 10^{40}$, it is 
unlikely but as mentioned in this limit of uncertainty.}}
%=========================================================================
\begin{table}[h]
\caption{Critical temperature ($T_{cl}$) of SQN's of different baryon number $b_N$, 
the total number $N_N$ of SQN's that coalesce together and their total final mass 
in solar mass units.}
\begin{tabular}{lcccr}
\hline
$b_N$ & $T_{cl}$ (MeV) & $N_N$ & $M/M_{\odot}$ \\
\hline
$10^{40}$ & $ 1.2$ & $2.40 \times 10 ^{10}$ & 0.42 \\
\hline
$10^{42}$ & $ 1.6$ & $2.44 \times 10 ^{14}$ & 0.24 \\
\hline
$10^{44}$ & $ 4.45$ & $1.23 \times 10 ^{11}$ & 0.01 \\
\hline
$10^{46}$ & $ 20.6 $ & $1.1 \times 10 ^{7}$ & 0.0001 \\
\hline
\end{tabular}
 
\label{tbl1}
\end{table}
%=======================================================================

It is clear however there can be no further clumping of those already 
clumped SQN's, the density of such objects would be too small within the 
horizon for further clumping. 

These objects will survive the entire evolutionary history of the universe 
uptil now. 

In recent years \cite{alcock93}, there has been experimental evidence 
for at least one form of dark matter - the Massive Astrophysical Compact 
Halo Objects (MACHO) detected through gravitational micro lensing 
~\cite{alcock93}. MACHOs only manifest gravity, thus detectable by 
gravitational micro lensing. These cosmic objects are dark, not visible 
astronomically ~\cite{raha05}. Based on 
\cite{raha05,alam99} Milky way halo MACHOs 
are detected in the direction of LMC-the Large Magellanic Cloud, 
MACHOs are expected to be in the mass range  (0.15-0.95) $M_{\odot}$, more 
likely to be in the vicinity of 0.5 $M_{\odot}$ \cite{alam99,bhattacharya00}, much higher 
than the fusion threshold of 0.08 $M_{\odot}$. For very good reasons MACHOs 
are unlikely candidates for white dwarfs , not even blue dwarfs. The 
suggestions \cite{ozel10} that they could be simply primordial black 
holes, can be ruled out since existence of primordial black holes need 
order of magnitude fluctuation, improbable in reality ! 

From all these considerations the clumped SQN's, still surviving from the 
primordial epoch of quark hadron phase transition in the universe are the
MACHO 's; again indicating that QN's with baryon number $b_N$ even less 
than $10^{42}$ is certainly a possibility, within the scope of the 
uncertainty of the parameters~\cite{raha05}. 

The relics of the first order cosmic phase transition from quarks to hadrons 
can thus lead to stable strange quark nuggets, candidates for cold dark matter, 
appearing in the form of MACHOs already experimentally observed. 

The very existence of SQN's in the form of MACHO's further substantiates 
the scenario of a first order 
phase transition leading to exotic SQN's. What is the structure of the 
distributions of SQN's in the universe? 

One possible scenario could be that the clumped SQN's will be attracted 
to each other by gravity, grow in size by devouring non strange 
objects and turning them strange, since energetically lower energy is 
the most plausible scenario. Such a scenario in extreme conditions has 
already been observed by the 
composite image, showing the ring like dark matter distribution superimposed 
on the optical view of galaxy cluster C10024X17\cite{nasa}
The other more probable scenario is that lumps of 
SQN's  are still floating around in our galaxy in no structured form. To our 
knowledge structured links of MACHOs are not known as yet.

\section{The Neutron Star}
The entire discussion so far, is related to very early universe when the 
baryonic chemical potential $\mu \rightarrow 0$ and the temperature is of the 
order of (150-200) MeV. The cosmological big bang is played out at LHC 
albeit in a miniature scale, with the little bangs between two nuclei. 
As is well known the big bang is a display of gravity, space and 
time where as the little bang is essentially to do with confinement and 
subsequently to deconfinement in extreme conditions. 

On the other extreme end of the phase diagramme lies a domain of very high 
baryonic density but at rather low temperature, a scenario for neutron 
star matter, of compressed baryonic matter and a temperature, very near 
zero.

It is widely conjectured \cite{ozel10,glendenning97,alford08} that the 
quark gluon sector of such 
matter may indeed consist of "colour super conductors" and high density 
hadronic (neutron) matter or hybrid matter in the hadronic sector\cite{alford08}. 

We study the spin down behaviour of a rotating neutron star with the 
realisation that changes in the internal structure as the star spins down, 
will be reflected in the moment of inertia and hence the deceleration. In 
this letter we are not considering the "recycling" scenario of binary 
system. 

During the spin down of a (say ) millisecond neutron star, the central 
\cite{glendenning97} density increases with decreasing centrifugal force; leading 
to a phase transition from the somewhat incompressible nuclear matter to 
the highly compressible, perfect fluid, quark matter in the stellar core. 

Indeed as the bulge of quark matter in the stellar core increases in 
dimension, a perfect fluid of QCD colour will set in, and, the perfect 
colour fluid will splash into hadronic matter transforming more of hadronic 
matter to colour superconducting quark matter. After the quark gluon matter 
dominates in the core, the star would contract significantly and its moment 
of inertia decreases sharply, a common signature of phase transition from 
confined to deconfined matter. 

Glendenning \cite{glendenning97} sometime ago pointed this phenomenon by introducing 
the braking index; for completeness we quote, 
$n(\Omega)=\dot{\Omega} \Omega/{\dot{\Omega}}^2=
3-\frac{3I'(\Omega) \Omega +I"(\Omega) {\Omega}^2}
{2I(\Omega)+I'(\Omega)\Omega} $, 
for a frequency of $\Omega$. The other notations have their usual meaning. 

During phase transition $n(\Omega)$ will deviate \cite{glendenning97} 
substantially from its canonical value $n=3$. Since the growth is paced by 
slow spin down of the pulsar, the signal of phase transition will be "on" 
over a long time-the slow increase in the entropy in the new phase (quark) 
will lead to slowing down of the spin. 

The nature of the phase transition from hadrons to quarks in a neutron star, 
thus is unique and very different; from the experiments carried out on our 
earth. The continuous process of phase transition closely resembles cross 
over but not exactly identical. It is felt that by means of 
designing ingenious experiments conducted by "CBM" type of detector this 
novel matter can be discovered; one possibility of course is to study "CBM" 
but at cooler environment, analogous to the core of neutron star.

By now a very large number and variety of neutron stars have been discovered 
\cite{ozel10,alford08}. Thus it is quite realistic to see the relationship of 
M (neutron star) and the radius R and try to extract the equation of state 
more precisely. For the internal quark structure we need a softer equation of 
state (quark matter being more compressible) and even more compact star. 
This is being explored in~\cite{sinha}. 

Recently ~\cite{demorset10,ozellett10} existence of the massive 
Pulsar PSRJ 1614-2230 
has been reported with mass of 1.97$\pm$0.04 $M_{\odot}$, the structure 
of the neutron star has gone through substantial rethinking. In particular, 
the highest mass neutron star that can be supported against collapse 
depends very sensitively on the underlying equation of state. In 
particular, if quark, hyperon or boson degrees of freedom are excited 
at high densities the equation of state softens and cannot support 
massive neutron star~\cite{alford08}. Even a single event of a massive 
neutron star can therefore strongly constrain the fundamental properties 
of ultradense matters. 

Following the phenomenological equation of state proposed by 
Alford {\it et al.} ~\cite{alford05}, {$\dot{O}$}zel {\it et al.} 
~\cite{ozellett10} demonstrated that strongly interacting quark core 
can sustain the large mass very high density inside the neutron star 
such a scenario is plausible. On the otherhand strongly interacting 
quark matter has already been discovered at RHIC and LHC, indicate the 
quark gluon plasma formed after the collision of two nuclei will not 
lead to non-interacting fermi gas but strongly interacting fluid. 
It has been suggested by the four RHIC experimental groups 
~\cite{rhicwhitepaper} that the best parameters for describing what they 
found in the system created in Au+Au collisions was one usually applied 
to liquids-namely, the ratio of $\eta/s$ for its shear viscosity to its 
entropy density. That ratio turned out to be nearly zero, making the 
system one of the first experimentally acessable "perfect fluid" ever 
observed in the laboratory.

It will be of some interest to compare and contrast the "perfect fluid" 
property of the quark matter in the microsecond universe with the 
"perfect fluid" of the core of the neutron star.  

It is interesting to note that for the early universe, we have a depleting 
quark matter as hadronisation progress and the universe expands in space and 
time. From the canonical value of $\eta/s \le 1/{4\pi}$, with hadronisation 
$\eta/s$ will go on increasing as pointed at Roy {\it et al.} 
\cite{royalacey07}. 

Eventually, the SQN's will be floating in a dilute hadronic fluid, which is not 
so perfect, facing more viscous drag than its its quark matter counter 
part.

In the case of the neutron star however the scenario is opposite, 
more hadrons will be transformed to quarks so $\eta/s$ will 
decrease towards the canonical value $\eta/s \le 1/{4\pi}$. 
For the neutron star however an approximate estimate of 
$\eta/s \sim T \lambda_F c_s$ will indicate that with very low value of 
$\lambda_F \equiv (\rho \sigma)^{-1}$ with high (very high $\rho$) and 
extremely low temperature $\eta/s$ for the quark core of the star with 
$c_s \sim 1/{\sqrt 3}$ (say) may well 
%In the case of neutron star however the sceario is exactly the opposite, 
%more hadrons will be transformed to quarks as the star spins down pushing 
%$\eta/s$ more towards the canonical value $\eta/s \le 1/{4\pi}$. For the 
%neutron star matter an approximate estimate $\eta/s \sim T \lambda_F c_s$ 
%will indicate for $c_s \equiv 1/{\sqrt{3}}$ (say) and very low value of 
%$\lambda_F=(\rho\sigma)^{-1}$ with very high $\rho$ and extremely low 
%temperature $\eta/s$ of quark core of the neutron star may 
go down  
below the generic value $1/{4\pi}$ and close to zero \cite{buchel09} or 
indeed go to zero making the core, 
really a perfect fluid splashing on the membrane \cite{witten84} of hybrid 
hadronic matter and quark core. The discovery 
of massive neutron star, 1.97 $\pm$ $M_{\odot}$ further substantiates 
this view point. It will be of great interest to explore this in future. 

{\bf Acknowledgment:}
The author wishes to thank Sibaji Raha, Partha Mazumder, J. Alam, Larry McLerran, T. Kodama, 
Horst Stoecker, Aninda Sinha and Jajati K. Nayak for stimulating discussion. 
The author also wishes to thank the Department of Atomic Energy of India for 
the Homi Bhabha Chair.

\end{document}